\documentclass[12pt]{article}

\usepackage{amsthm,amsfonts,amssymb,amscd,mathrsfs,physics,bm}
\usepackage{amsfonts}
\usepackage{amssymb}
\usepackage{color}
\usepackage{epsfig}
\usepackage{geometry}

\reversemarginpar

\numberwithin{equation}{section}

\newtheorem{thm}{THEOREM}[section]

\newtheorem{rem}[thm]{Remark}

\DeclareMathOperator{\diag}{diag}

\begin{document}

\title{On a Center-of-Mass System of Coordinates for Symmetric Classical and Quantum Many-Body Problems}
\date{}

\author{\'{E}rik Amorim}

\maketitle

\begin{center}
  Department of Mathematics, Rutgers University \\
  Hill Center - Busch Campus\\
  110 Frelinghuysen Road\\
  Piscataway, NJ 08854-8019, USA  \\
  \texttt{erik.amorim@math.rutgers.edu} \\

\textrm{\small Version of June 26, 2019. Typeset with \LaTeX . }

\end{center}

\bigskip

\textrm{\small \textbf{Keywords:} Jacobi coordinates, center-of-mass system of coordinates, permutation symmetry}

\bigskip


\thispagestyle{empty}


\begin{abstract}
{\noindent 
In the context of classical or quantum many-body problems involving identical bodies, a linear change of coordinates can be constructed with the properties that it includes the center-of-mass as one of the new coordinates and preserves the inherent permutation symmetry of both the Hamiltonian and the admissible states. This has advantages over the usual system of Jacobi coordinates in the study of many-body problems for which permutation symmetry of the bodies plays an important role. This paper contains the details of the construction of this system and the proof that these properties uniquely determine it, up to trivial modifications. Examples of applications to both classical and quantum problems are explored, including a generalization to problems involving groups of different species of bodies.
 }
\end{abstract}

\vfill

\copyright(2019) \small{The author. Reproduction of this preprint, in its entirety, is permitted for non-commercial purposes only.}

\section{Introduction}

In non-relativistic classical and quantum $N$-body problems with a translation-invariant
Hamiltonian
\begin{equation}
 H = \sum_{1\leq j\leq N} \frac{p_j^2}{2m_j} + \mathop{\sum\sum}_{1\leq j< k\leq N} V_{jk}(|q_j - q_k|) \ ,
\end{equation}
where $q_j\in \mathbb{R}^3$ denotes the position vector of particle $j$ in some arbitrary
Galilei frame, $p_j$ its momentum, and $m_j$ its mass, the motion of the center-of-mass has no objective physical significance.
 Objectively significant are only the intrinsic properties of the $N$-body system.
 If $Q\in \mathbb{R}^3$ denotes the position vector of the center-of-mass in the same Galilei frame,
and $P$ its momentum, then the Galilei transformation
$q_j\mapsto r_j := q_j- Q$ and $p_j\mapsto \pi_j := p_j - \frac{m_j}{M} P$ (where $M$ is the total mass in the system) separates off the kinetic energy assigned to the center-of-mass, i.e. it
accomplishes
\begin{equation}
 H = \frac{P^2}{2 M} + \tilde{H} \ ,
\end{equation}
where
\begin{equation}
\tilde{H}
=
\sum_{1\leq j\leq N} \frac{\pi_j^2}{2m_j} + \mathop{\sum\sum}_{1\leq j< k\leq N} V_{jk}(|r_j - r_k|),
\end{equation}
is the ``intrinsic Hamiltonian" of the $N$-body system, encoding all the physically objective features of the $N$-body system. The $N$ position variables $r_j$ and the $N$ momentum variables $\pi_j$
are not independent but satisfy the center-of-mass frame constraints
\begin{equation}
\sum_{1\leq j\leq N} r_j = 0 \quad , \quad \sum_{1\leq j\leq N} \pi_j = 0 \ .
\end{equation}
Thus, in terms of the available degrees of freedom, the intrinsic $N$-body Hamiltonian is
actually equivalent to a non-translation-invariant $(N-1)$-body problem. Therefore it is desirable to find a transformation to new coordinates which expresses
$\tilde{H}$ as a function of $N-1$ independent positions and momenta, which can be thought of as pertaining to ``virtual bodies''. The so-called \emph{Jacobi coordinates} accomplish this feat (see section 2).
 
Now, systems whose bodies are identical (or systems involving different groups of identical bodies) enjoy valuable permutation symmetry or anti-symmetry properties in both their Hamiltonian and admissible states, which can play a determining role in their study. But it turns out that, after reducing such systems to $(N-1)$-body systems by employing Jacobi coordinates, one finds that they lose their symmetry properties and can no longer be studied by means of the same techniques. The goal of this paper is to build a center-of-mass coordinate change that preserves symmetry in whatever sense is meaningful for the problem at hand. We will prove that there is an essentially unique coordinate change with this property.

In the next section we give an example of a quantum system exhibiting symmetry, and we show how the Jacobi coordinates change is carried out and destroys this symmetry. Section 3 then shows the construction of our \emph{symmetric center-of-mass system of coordinates} applied to that same problem, proving its uniqueness in the process. Section 4 shows that the same system of coordinates is suitable also for symmetric classical problems, even though they involve fundamentally different notions of configuration space, admissible states and Hamiltonian. The main result is summarized in theorems (\ref{theorem}) and (\ref{thm_classical}), and the coordinate system in its most compact form in equations (\ref{CMsystem}) and (\ref{CMinverse}). Finally in the last section we indicate how to generalize the construction to problems involving more than one group of objects of the same type (uniqueness does not hold anymore), culminating in the change of coordinates described in theorem (\ref{thm_multi}).

\section{Jacobi coordinates: a quantum example}

To illustrate the need for a symmetric center-of-mass coordinate system, we start by discussing the model that inspired us to create it, which can be found in \cite{hogr},\cite{miki1},\cite{miki2}. It is a study of the asymptotic properties of the equilibrium configuration energy and ground state of a ``bosonic atom'' consisting of one positively charged nucleus and $N$ negatively-charged bosons as $N$ goes to infinity, assuming the so-called \emph{Born-Oppenheimer (BO) approximation}, that is, the nucleus is considered to sit immovable at the origin. Our motivation for introducing our system of coordinates was the desire to study the same system, by adapting the same techniques, but without the BO approximation.

But we emphasize that neither the particular type of interaction between the bodies in this model (Coulomb attraction/repulsion) nor the fact that they are bosons instead of fermions is what justifies the need for such a system; the important feature is the symmetry that comes from the fact that all (or all but one) of the bodies are identical.

Consider a quantum-mechanical system consisting of one distinguished particle of mass $m_0$ and charge $Z>0$, and $N$ identical particles of mass $m$ and charge $z<0$, all of which attract or repel each other via the Coulomb potential. The state of the system is given by a $\mathbb{C}$-valued (we ignore spin) wavefunction
\begin{equation}
L^2(\mathbb{R}^{3(1+N)})\ni\psi = \psi(q_0,q_1,\ldots,q_N) \quad , \quad q_j\in\mathbb{R}^3, j=0,1,\ldots,N
\end{equation}
where $q_0$ is the position of the zeroth particle and each $q_j$, $j\geq 1$, is the position of one of the other particles. Born's Rule says that $|\psi(q_0,q_1,\ldots,q_N)|^2$ is the probability density for the zeroth particle to be at $q_0$ and for there to exist one of the other particles at each $q_1,\ldots,q_N$. Because of indistinguishability, the only admissible wavefunctions are those that satisfy the \emph{symmetry condition}\footnote{The reason why this condition must be stated in this form, and not in the weaker form $|\psi(q_0,q_1,\ldots,q_N)|^2 = |\psi(q_0,q_{\sigma(1)},\ldots,q_{\sigma(N)})|^2$ that one might expect from the Born rule, is that it should be true that the expected value of any observable (self-adjoint operator) $A$, that is, the inner product $\langle \psi,A\psi\rangle_{L^2}$, should be independent of permutations of any but the zeroth variable. Since quantum-mechanical observables are not restricted to simple multiplication operators, but rather can take the form of differential operators as well, it turns out that the condition needed is the one given in the text. See \cite{griff} for details about this and the Pauli exclusion principle for the case of fermions.}
\begin{equation}
\psi(q_0,q_1,\ldots,q_N) = \psi(q_0,q_{\sigma(1)},\ldots,q_{\sigma(N)}) \quad \text{for all permutations } \sigma\in S^N
\end{equation}
if particles 1 through $N$ are bosons, or the \emph{anti-symmetry condition} (Pauli exclusion principle)
\begin{equation}
\psi(q_0,q_1,\ldots,q_N) = \mathop{sgn}(\sigma)\psi(q_0,q_{\sigma(1)},\ldots,q_{\sigma(N)}) \quad \text{for all permutations } \sigma\in S^N
\end{equation}
if they are fermions ($S^N$ denotes the symmetric group in $N$ objects).

The Hamiltonian operator, defined only for twice-differentiable wavefunctions (but it does admit a self-adjoint extension to a larger domain - see \cite{lieb}, \cite{ls}, \cite{rs}), is given by summing the kinetic energy differential operators of each particle and the potential energy multiplication operators of each pair of particles:
\begin{equation}
\label{H}
H = -\frac{\hbar^2}{2m_0} \Delta_0 -\frac{\hbar^2}{2m}\sum_{j=1}^N \Delta_j - Zz\sum_{j=1}^N \frac{1}{|q_j-q_0|} + z^2\mathop{\sum_{i=1}^N\sum_{j=1}^N}_{i<j} \frac{1}{|q_i-q_j|} \ .
\end{equation}
Here the notation $\Delta_j$ indicates the Laplacian operator acting only with respect to $q_j\in\mathbb{R}^3$, that is,
\begin{equation}
\Delta_j = \pdv[2]{}{x_j} + \pdv[2]{}{y_j} + \pdv[2]{}{z_j} \quad , \quad \text{where } q_j = (x_j,y_j,z_j).
\end{equation}
Associated to the operator $H$ is the quadratic functional that yields the expected value of the energy in the state $\psi$, obtained by formally computing the $L^2(\mathbb{R}^{3(1+N)})$ inner product $\langle \psi,H\psi\rangle$ with the help of an integration by parts in the kinetic terms:
\begin{multline}
Q(\psi) = \frac{\hbar^2}{2m_0}\int |\nabla_0\psi|^2 + \frac{\hbar^2}{2m}\sum_{j=1}^N \int |\nabla_j\psi|^2 - Zz\sum_{j=1}^N \int \frac{|\psi|^2}{|q_j-q_0|} + \\ + z^2\mathop{\sum_{i=1}^N\sum_{j=1}^N}_{i<j} \int \frac{|\psi|^2}{|q_i-q_j|}
\label{Q}
\end{multline}
with the analogous remark about the notation $\nabla_j$ as for the Laplacian above. It turns out that this functional is bounded below when computed on $H^1$ functions of $L^2$ norm $1$ (for details see \cite{lieb},\cite{rs}), and the infimum is interpreted as the equilibrium energy of the system.

\begin{rem} The (anti-)symmetry of $\psi$ permits us to make statements such as
\begin{equation}
\int |\nabla_j\psi|^2 = \int |\nabla_1\psi|^2 \; , \;
\int\frac{|\psi|^2}{|q_j-q_0|} = \int \frac{|\psi|^2}{|q_1-q_0|} \; , \;
\int \frac{|\psi|^2}{|q_i-q_j|} = \int \frac{|\psi|^2}{|q_1-q_2|}
\label{symmetry}
\end{equation}
for all $i\neq j$, which combined with the symmetry of $H$ are useful in Hartree and Hartree-Fock theory for studying asymptotic properties of the equilibrium energy, because they allow the replacement of all indices $i$ and $j$ in $Q$ by 1 and 2, thus re-expressing $Q$ in terms of a conditional two-body functional (depending only on variables $q_1,q_2$ and conditioned on $q_0$). For details see \cite{miki2} or \cite{roug}.
\end{rem}

However, it is easy to argue that there will not exist a \emph{ground-state}: the infimum can never be attained. Indeed, by plugging in trial functions of the form $\psi = \psi_0(q_0)\phi(q_1,\ldots,q_N)$, we see that $\psi_0$ will only contribute to the first term of the energy $Q(\psi)$ (the variable $q_0$ in the first potential term disappears after a translation change of variables in the integral), which can be made arbitrarily small by reducing $\int |\nabla\psi_0|^2$, but never zero because $\psi_0$ must have positive $L^2$ norm\footnote{If this were a classical problem instead, where it \emph{is} possible to reduce the contribution of the zeroth kinetic energy to zero, then it is also easy to see that there would not exist a \emph{unique} minimizer, because the problem is translation-invariant.}. On the other hand, the functional $Q$ contains more than the portion of the energy that we are interested in, because contained in the kinetic energy part is the energy of motion of the system as a whole (the kinetic energy of the center-of-mass). But there doesn't exist an operator associated to the kinetic energy of the system that can be subtracted from $H$ in order to isolate the interesting part; the way to achieve this separation is to first change coordinates into a system which includes the center-of-mass as a coordinate:
\begin{equation}
T:(q_0,q_1,\ldots,q_N) \mapsto (\xi_0,\xi_1,\ldots,\xi_N) \quad , \quad \xi_0 = \frac{1}{M_{\mathrm{Tot}}}(m_0q_0+mq_1+\ldots+mq_N)
\end{equation}
(we abbreviated the total mass $m_0+Nm$ with $M_{\mathrm{Tot}}$) and write $\psi$ as an $L^2$-normalized function of the new $\xi$ coordinates:
\begin{equation}
\chi(\xi_0,\xi_1,\ldots,\xi_N) = |\det T|^{-1/2}\psi(T^{-1}(\xi_0,\xi_1,\ldots,\xi_N)) \ ,
\label{new_wavefunction}
\end{equation}
then finally express all terms in (\ref{H}) using $\chi$ instead of $\psi$, obtaining a new Hamiltonian $\tilde{H}$ such that
\begin{equation}
H\psi = \tilde{H}\chi
\label{H_tilde}
\end{equation}
whenever $\psi$ and $\chi$ are related by (\ref{new_wavefunction}). If the coordinate change $T$ is linear, one of the terms in $\tilde{H}$ will involve the Laplacian with respect to $\xi_0$ (as will become clear in (\ref{cross_terms_computation}) below), and throwing out this term will leave us with an operator associated to the desired energy of the system.

This process might end up introducing unhelpful \emph{cross-terms}: the Chain Rule applied to (\ref{new_wavefunction}) gives (with $\Delta\psi = \nabla\cdot\nabla\psi$)
\begin{equation}
\Delta_j\psi(\bm{q}) = |\det T|^{1/2}\sum_{k,l=0}^N \pdv{\xi_k}{q_j}\pdv{\xi_l}{q_j}\nabla_k\cdot\nabla_l\chi(T\bm{q}) \ .
\end{equation}
We call \emph{cross-term} an expression of the form $k$-divergence of $l$-gradient for $k\neq l$, of which there are none in the original Hamiltonian $H$. A commonly employed family of coordinate changes that prevent the appearance of such cross-terms can be found for example in \cite{fo}, \cite{gim} and \cite{post}, usually referred to by the name \emph{Jacobi coordinates}. It is well-known that the crucial property needed to preclude cross-terms is orthogonality of the matrix $\displaystyle\pdv{\bm{q}}{\bm{\xi}}$ (after suitable rescalings to make all masses equal to 1 - see the next section). It is also well-known that one can construct such matrices even when the objects in the system have different masses - see remark (\ref{remark}) below. One possible instance of a Jacobi coordinate change consists in starting with the separation between two of the bodies as a new coordinate, then iteratively constructing the others as the separation between the next body and the center-of-mass of the previously used bodies (different normalizing scale factors can be included too). For our problem, then, a Jacobi system of coordinates could look like
\begin{equation}
\left\{\begin{array}{lll}
\xi_0 & = & (m_0q_0+mq_1+\ldots+mq_N)/M_{\mathrm{Tot}} \\
\xi_1 & = & q_1-q_2 \\
\xi_2 & = & (q_1+q_2)/2 - q_3 \\
\vdots & & \vdots \\
\xi_{N-1} & = & (q_1+q_2+\cdots+q_{N-1})/(N-1) - q_N \\
\xi_{N} & = & (q_1+q_2+\cdots+q_{N})/N - q_0
\end{array}\right.
\label{Jacobi}
\end{equation}
Employing this system for $N=1$ (a two-body problem), what is obtained after (\ref{H_tilde}) is
\begin{equation}
\tilde{H} = -\frac{\hbar^2}{2(m_0+m)}\Delta_0 - \frac{\hbar^2}{2\mu}\Delta_1 + \frac{Zz}{|\xi_1|}
\end{equation}
where
\begin{equation}
\mu = \frac{Mm}{M+m}
\label{2D_Jacobi}
\end{equation}
is called the \emph{reduced mass}. Throwing away the first term of $\tilde{H}$ gives us the Hamiltonian for a one-body problem with mass $\mu$ in a central potential, known as the \emph{Kepler problem}. It admits a ground state energy and a unique ground state configuration conditioned on the position $\xi_0$ of the center-of-mass.

But for $N>1$ the symmetry of the potential part of the Hamiltonian is hopelessly lost under the change (\ref{Jacobi}), because one can compute and check that $|\xi_i-\xi_j| \neq |\xi_k-\xi_l|$ if $\{i,j\} \neq \{k,l\}$. Further, the symmetry or anti-symmetry condition on the wavefunction $\psi$ does not translate to anything practical about permutation of the variables $\xi_1,\ldots,\xi_N$ of $\chi$. If we want to study the properties of $\tilde{H}$ using the same techniques as one would for the symmetric $H$ and its (anti-)symmetric wavefunctions, a better change of coordinates is clearly needed.

\section{Symmetric center-of-mass coordinates (quantum case)}

Here we describe our coordinate system, illustrated with the same system as in the previous section, and explain in which sense and under which conditions it is unique.

The coordinate change should be an invertible map $T:\mathbb{R}^{3(1+N)}\to\mathbb{R}^{3(1+N)}$ for which we require the following conditions:
\begin{itemize}
\item[1.] linearity and independence from Cartesian coordinates;
\item[2.] one of the new coordinates is the center-of-mass;
\item[3.] the structure of Hamiltonian is preserved and includes $-(\hbar^2/2M_{\mathrm{Tot}})\Delta_0$, the kinetic energy operator of the system in the new coordinates;
\item[4.] symmetry of wavefunctions is preserved.
\end{itemize}
Let us elaborate on each and see how they restrict the possible transformations $T$ further and further.

\smallskip
\noindent\textbf{Condition 1.} We demand linearity for simplicity of computation and to avoid singularities. But in order to avoid bringing to the fore such physically meaningless quantities as the scalar coordinates of each particle position, we look for a linear transformation that operates only on the level of the positions of each particle, that is, only on the vectors $q_0,q_1,\ldots,q_N$ as opposed to explicitly referencing the $x,y,z$ coordinates of the particles. So we think of $\mathbb{R}^{3(1+N)}$ in both its domain and range as $(\mathbb{R}^3)^{1+N}$, and we write $T(q_0,q_1,\ldots,q_N) = (\xi_0,\xi_1,\ldots,\xi_N)$, each $q_j$ and $\xi_j$ in $\mathbb{R}^3$. This can be made more formal by saying that $T$ is a tensor product of a $(1+N)\times(1+N)$ matrix $\tilde{T}$ with the $3\times 3$ identity matrix. But to avoid cluttered notation, we denote the elements of $\tilde{T}$ by $T_{ij}$ ($0\leq i,j \leq N$), without tildes and hopefully without confusion.

\smallskip
\noindent\textbf{Condition 2.} We impose that $\xi_0$ be the center-of-mass of the system:
\begin{equation}
\xi_0 = \frac{1}{M_{\mathrm{Tot}}}(m_0q_0+mq_1+\cdots+mq_N) \ .
\label{center_of_mass}
\end{equation}
With this the first row of the matrix of $\tilde{T}$ is already determined.

\smallskip
\noindent\textbf{Condition 3.} We want $T$ to transform the structure of the Hamiltonian into a form similar to $H$: kinetic plus potential terms, with the kinetic terms of identical particles appearing with equal weights, the same being true of the potential terms of interaction between similar pairs. Additionally, what should sit in front of the $\Delta_0$ term is the fraction $-\hbar^2/2M_{\mathrm{Tot}}$, so that this term becomes the kinetic energy operator of the whole system. Let us only worry about the kinetic term in (\ref{H}) and later study what happens to the potential terms. We stipulate that there should exist some constant $\mu>0$ (which we call the \emph{reduced mass}) such that
\begin{equation}
-\frac{\hbar^2}{2m_0}\Delta_0\psi(\mathbf{q}) - \frac{\hbar^2}{2m} \sum_{j=1}^N \Delta_j\psi(\mathbf{q}) = -\frac{\hbar^2}{2M_{\mathrm{Tot}}}\Delta_0\chi(\bm{\xi}) - \frac{\hbar^2}{2\mu}\sum_{j=1}^N \Delta_j\chi(\bm{\xi}) \ ,
\label{kinetic}
\end{equation}
where $\bm{\xi} = T\mathbf{\bm{q}}$ and $\chi$ is defined by
\begin{equation}
\chi(\mathbf{\bm{\xi}}) = |\det T|^{-1/2} \psi(T^{-1}\mathbf{\bm{\xi}}) \quad , \quad \bm{\xi}\in\mathbb{R}^{3(1+N)} \ .
\label{new_wavefunction_again}
\end{equation}
By the Chain Rule,
\begin{equation}
\nabla_j\psi(\bm{q}) = |\det T|^{1/2} \sum_{k=0}^N T_{kj} \nabla_k\chi(T\bm{q})
\end{equation}
and
\begin{equation}
\Delta_j\psi(\bm{q}) = (\det T)^{1/2} \sum_{k,l=0}^N T_{kj} T_{lj} \nabla_k \cdot\nabla_l\chi(T\bm{q}) \ .
\end{equation}
Omitting the $\bm{q}$ argument of $\psi$ and the $T\bm{q}$ argument of $\chi$, we then have
\begin{equation}
\begin{split}
-\frac{\hbar^2}{2m_0}&\Delta_0\psi - \frac{\hbar^2}{2m}\sum_{j=1}^N \Delta_j\psi \\
&= -|\det T|^{1/2} \left( \frac{\hbar^2}{2m_0}\sum_{k,l=0}^N T_{k0} T_{l0} \nabla_k\cdot\nabla_l\chi + \frac{\hbar^2}{2m}\sum_{j=1}^N\sum_{k,l=0}^N T_{kj} T_{lj} \nabla_k\cdot\nabla_l\chi \right) \\
&= -|\det T|^{1/2} \frac{\hbar^2}{2} \sum_{k,l=0}^N \left( \frac{1}{m_0} T_{k0} T_{l0} + \frac{1}{m}\sum_{j=1}^N T_{kj} T_{lj} \right) \nabla_k\cdot\nabla_l\chi \ ,
\end{split}
\label{cross_terms_computation}
\end{equation}
so that we can achieve (\ref{kinetic}) by imposing
\begin{equation}
|\det T|^{1/2}\left(\frac{1}{m_0} T_{k0} T_{l0} + \frac{1}{m}\sum_{j=1}^N T_{kj} T_{lj} \right) = \begin{cases}
1/M_{\mathrm{Tot}} & , \;\;k=l=0 \ , \\
1/\mu & , \;\;k=l>0 \ , \\
0 & , \;\;k\neq l \ .
\end{cases}
\label{orthogonality}
\end{equation}
Given our choice of $\xi_0$, the condition for $k=l=0$ holds if and only if
\begin{equation}
\det T = \pm 1 \ .
\end{equation}
Now rewrite property (\ref{orthogonality}) as
\begin{equation}
\tilde{T} R^{-1} \tilde{T}^t = S^{-1}
\label{TRS}
\end{equation}
where the $(1+N)\times(1+N)$ matrices $R$ and $S$ are given by
\begin{equation}
R = \diag(m_0,m,\ldots,m) \qquad , \qquad S = \diag(M_{\mathrm{Tot}},\mu,\ldots,\mu) \ .
\label{RS}
\end{equation}
Then
\begin{equation}
(\det T)^2 = \frac{\det(R)}{\det(S)} \quad \Longrightarrow \quad \det T = \pm\sqrt{\frac{m_0m^N}{M_{\mathrm{Tot}} \mu^N}}
\end{equation}
and since we need $|\det T|=1$, we can find $\mu$:
\begin{equation}
\mu = m\left(\frac{m_0}{M_{\mathrm{Tot}}}\right)^{1/N}.
\end{equation}

\begin{rem} 
\label{remark}
\label{differnet_masses}
If particles $1,\ldots,N$ in the system were not identical and had potentially different masses $m_1,\ldots,m_N$, as is the case in various examples of many-body problems, then condition 4 (as elaborated below) and the preservation of the symmetry of the potential energy would be meaningless; however it could still be desirable to find a center-of-mass system of coordinates satisfying conditions 1 and 2 that also prevents the appearance of unwieldy cross-terms in the transformed Hamiltonian. In this case one would stipulate the condition
\begin{equation}
    -\frac{\hbar^2}{2m_0}\Delta_0\psi(\mathbf{q}) - \frac{\hbar^2}{2} \sum_{j=1}^N \frac{1}{m_j} \Delta_j\psi(\mathbf{q}) = -\frac{\hbar^2}{2M_{\mathrm{Tot}}}\Delta_0\chi(\bm{\xi}) - \frac{\hbar^2}{2}\sum_{j=1}^N \frac{1}{\mu_j} \Delta_j\chi(\bm{\xi}) \ ,
\end{equation}
for some numbers $\mu_1,\ldots,\mu_N>0$. Proceeding as in the computations above, one would find
\begin{equation}
    \tilde{T} \cdot \mathrm{diag}(m_0,m_1,\ldots,m_N)^{-1} \cdot \tilde{T}^t = \mathrm{diag}(M_{\mathrm{Tot}},\mu_1,\ldots,\mu_N)^{-1} \ .
\end{equation}
This is easily achieved: choose an orthogonal matrix $\mathcal{O}$ whose zeroth row is given by
\begin{equation}
    (\mathcal{O}_{0j})_{j=0,1,\ldots,N} = \left( \sqrt{\frac{m_0}{M_{\mathrm{Tot}}}} , \sqrt{\frac{m_1}{M_{\mathrm{Tot}}}} ,\ldots, \sqrt{\frac{m_N}{M_{\mathrm{Tot}}}} \right)
\end{equation}
so that condition 2 is met, and let
\begin{equation}
    \tilde{T} = \mathrm{diag}\left(\frac{1}{\sqrt{M_{\mathrm{Tot}}}},\frac{1}{\sqrt{\mu_1}},\ldots,\frac{1}{\sqrt{\mu_N}}\right) \cdot \mathcal{O} \cdot \mathrm{diag}(\sqrt{m_0},\sqrt{m_1},\ldots,\sqrt{m_N}) \ .
    \label{O}
\end{equation}
The only additional constraint comes from $|\det \tilde{T}| = 1$, which, when implemented in (\ref{O}) above, produces a restriction on the possible choices of $\mu_j$'s:
\begin{equation}
    m_0m_1\cdots m_N = M_{\mathrm{Tot}}\mu_1\cdots\mu_N \ .
\end{equation}
\end{rem}

\smallskip
\noindent\textbf{Condition 4.} $T$ must have the property that, if $\psi$ is (anti-)symmetric in the variables $q_1,\ldots,q_N$, then $\chi$ defined as in (\ref{new_wavefunction_again}) is (anti-)symmetric in $\xi_1,\ldots,\xi_N$ as well. Let us work out what this implies.

For a given permutation $\sigma\in S^N$ let $\sigma$ also denote the isomorphism
\begin{equation}
\sigma: \mathbb{R}^{3(1+N)} \to \mathbb{R}^{3(1+N)} \quad , \quad \sigma(x_0,x_1,\ldots,x_N) = (x_0,x_{\sigma(1)},\ldots,x_{\sigma(N)})
\end{equation}
(each $x_j\in\mathbb{R}^3$). The required equality $\chi = \chi\circ \sigma$ (symmetric case), which translates to
\begin{equation}
|\det T|^{-1/2} \psi(T^{-1}\bm{\xi}) = \chi(\bm{\xi}) = \chi(\sigma \bm{\xi}) = |\det T|^{-1/2} \psi(T^{-1}\sigma\bm{\xi}) \ ,
\label{chain}
\end{equation}
holds for any wavefunction $\psi$ symmetric in all but the zeroth variables, for all $\sigma\in S^N$ and all $\xi\in\mathbb{R}^{3(1+N)}$, if and only if to every $\sigma\in S^N$ corresponds $\pi\in S^N$ such that
\begin{equation}
\pi T^{-1} = T^{-1}\sigma
\label{T_symm}
\end{equation}
(simply compare the arguments of $\psi$ on both ends of (\ref{chain}) and use symmetry of $\psi$). In the case of anti-symmetry, for fermionic particles, $\pi$ must also have the same sign as $\sigma$. We remind the reader that $T=\tilde{T}\otimes I_{3\times 3}$ for some $\tilde{T}:\mathbb{R}^{1+N}\to\mathbb{R}^{1+N}$, and of course any $\sigma\in S^N$ acts on $\mathbb{R}^{1+N}$ by permuting coordinates 1 through $N$ with respect to the canonical basis $\{e_0,e_1,\ldots,e_N\}\subseteq\mathbb{R}^{1+N}$. So let
\begin{equation}
\tilde{T}e_1 = \sum_{j=0}^N a_j e_j \quad , \quad a_j\in\mathbb{R} \ .
\label{e_1}
\end{equation}
Then condition (\ref{T_symm}), after multiplying on left and right by $T$, means, for any $\sigma$, that $\sigma\cdot\tilde{T}e_1$ must be equal to one of the vectors
\begin{equation}
\tilde{T}e_1,\ldots,\tilde{T}e_N \ .
\label{vectors}
\end{equation}
But acting with $\sigma$ on $\tilde{T}e_1$ has the effect of permuting the coefficients $a_j$ for $j>0$, and the total number of different permutations that can be formed is
\begin{equation}
\frac{N!}{n_1!\cdots n_m!}
\label{comb}
\end{equation}
where $n_1,\ldots,n_m$ are the cardinalities of each set of repeated coefficients among the $a_j$'s, $j=1,\ldots,N$, with $n_1+\cdots+n_m=N$. Since the number in (\ref{comb}) must be equal to the number $N$ of different vectors in the list (\ref{vectors}), we need
\begin{equation}
(N-1)! = n_1!\cdots n_m!
\end{equation}
which can only happen\footnote{To see this, consider what happens when $N-1$ is prime.} for all $N$ if $m=2$ and one of the $n_j$'s is $N-1$ and the other $1$. Therefore $\tilde{T}e_1$ written in the basis $\{e_j\}$ must have $N-1$ repeated coefficients among the ones from $1$ to $N$. Analogously the same is true of each $\tilde{T}e_j$ for $j=2,\ldots,N$, and moreover they all share the same zeroth coefficient. Also note that (\ref{T_symm}) implies that performing any permutation on the rows of $T$ (except the zeroth) should yield the same as performing it on the columns instead; hence the coefficients on the zeroth column of $\tilde{T}$ are also all equal, except the zeroth. Finally, by relabeling the nonzero indices (that is, by replacing $T$ with $T\circ \sigma$ for an appropriate $\sigma$), and also remembering (\ref{center_of_mass}), we arrive at a matrix of the form
\begin{equation}
T = \begin{pmatrix}
\displaystyle\frac{m_0}{M_{\mathrm{Tot}}} & \displaystyle\frac{m}{M_{\mathrm{Tot}}} & \displaystyle\frac{m}{M_{\mathrm{Tot}}} & \displaystyle\frac{m}{M_{\mathrm{Tot}}} & \ldots & \displaystyle\frac{m}{M_{\mathrm{Tot}}} \\
C & A & B & B & \ldots & B \\
C & B & A & B & \ldots & B \\
C & B & B & A & \ldots & B \\
\vdots & \vdots & \vdots & \vdots & \ddots & \vdots \\
C & B & B & B & \ldots & A \\
\end{pmatrix}_{(1+N)\times(1+N)} \otimes I_{3\times 3}
\label{T_matrix}
\end{equation}
for some constants $A,B,C$ to be determined.

We refer back to (\ref{orthogonality}). The cases $k=l>0$, $0\neq k\neq l\neq 0$ and $0=k\neq l$, in that order, give equations for the entries $A,B,C$ in (\ref{T_matrix}):
\begin{equation}
\frac{C^2}{m_0} + \frac{A^2}{m} + \frac{(N-1)B^2}{m} = \frac{1}{\mu} \quad \Longrightarrow \quad A^2 + (N-1)B^2 = \frac{m}{\mu} - \frac{mC^2}{m_0} \ ,
\label{system_1}
\end{equation}
\begin{equation}
\frac{C^2}{m_0} + \frac{2AB}{m} + \frac{(N-2)B^2}{m} = 0 \quad \Longrightarrow \quad 2AB + (N-2)B^2 = -\frac{mC^2}{m_0} \ ,
\label{system_2}
\end{equation}
\begin{equation}
C + A + (N-1)B = 0 \quad \Longrightarrow \quad A + (N-1)B = -C \ .
\label{system_3}
\end{equation}
Subtract (\ref{system_2}) from (\ref{system_1}) and write $A^2 - 2AB + B^2$ as $(A-B)^2$ to get an expression for $A$ in terms of $B$. Plug that into (\ref{system_3}) to find $B$ in terms of $C$:
\begin{equation}
A = B \stackrel{(1)}{\pm} \sqrt{\frac{m}{\mu}} \quad , \quad B = -\frac{1}{N}\left( C \stackrel{(1)}{\pm} \sqrt{\frac{m}{\mu}} \right) \ .
\label{pm}
\end{equation}
The choice of the $\pm$ sign has to be the same in these two expressions, and that's what the $(1)$ above them signifies. With these, (\ref{system_1}) becomes
\begin{equation}
\frac{C^2}{N} - \frac{m}{N\mu} = -\frac{mC^2}{m_0} \ ,
\end{equation}
which can be solved to yield
\begin{equation}
C \ = \ \stackrel{(2)}{\pm}\sqrt{\frac{m_0m}{M_{\mathrm{Tot}}\mu}} \ .
\label{C}
\end{equation}
This $\pm$ sign has nothing to do with the choice of $\pm$ in (\ref{pm}), and we keep track of that with the $(2)$ above it. Now it becomes convenient to rewrite $A,B,C$ in terms of only the constant $r=m_0/M_{\mathrm{Tot}} = (\mu/m)^N$:
\begin{equation}
A = B \stackrel{(1)}{\pm} r^{-\frac{1}{2N}} \quad , \quad B = -\frac{1}{N}( \stackrel{(2)}{\pm} r^{\frac{1}{2}-\frac{1}{2N}} \stackrel{(1)}{\pm} r^{-\frac{1}{2N}} ) \quad , \quad C = \stackrel{(2)}{\pm} r^{\frac{1}{2}-\frac{1}{2N}} \ .
\label{ABC}
\end{equation}
We can also write the new coordinates compactly by using the \emph{empirical average}
\begin{equation}
\overline{\bm{q}} = (q_1+\cdots+q_N)/N \ .
\label{empirical_average}
\end{equation}
Computing $\xi_1$ (a similar computation gives $\xi_j$), we see how $\overline{\bm{q}}$ shows up due to the relationship between $A$ and $B$:
\begin{equation}
\begin{split}
\xi_1 &= Cq_0 + Aq_1 + B(q_2+\cdots+q_N) \\
&= Cq_0 + B(q_1+\cdots+q_N) \stackrel{(1)}{\pm} r^{-\frac{1}{2N}} q_1 \\
&= Cq_0 + (NB)\overline{\bm{q}} \stackrel{(1)}{\pm} r^{-\frac{1}{2N}} q_1 \ .
\end{split}
\end{equation}
Substituting the values of $A,B,C$, we obtain the final expression of our coordinate change:
\begin{equation}
\left\{\begin{array}{ccl}
\xi_0 & = & (m_0q_0+mq_1+\ldots+mq_N)/M_{\mathrm{Tot}} \ , \\
\xi_j & = & r^{-\frac{1}{2N}}\left( \stackrel{(1)}{\pm}(q_j-\overline{\bm{q}}) \stackrel{(2)}{\pm}\sqrt{r}(q_0-\overline{\bm{q}}) \right) \quad , \quad j=1,\ldots,N \ .
\end{array}\right.
\label{CMsystem}
\end{equation}
We have thus proved the following:
\begin{thm} For given $m_0,m>0$ and up to two independent choices of $\pm$ signs and relabeling of nonzero indices, the only family of transformations $T:\mathbb{R}^{3(1+N)}\to\mathbb{R}^{3(1+N)}$ (indexed by $N$) satisfying the following:
\begin{itemize}
\item[1.] $T=\tilde{T}\otimes I_{3\times 3}$ for some linear isomorphism $\tilde{T}:\mathbb{R}^{1+N}\to\mathbb{R}^{1+N}$;
\item[2.] the zeroth component of $T(q_0,q_1,\ldots,q_N)$ in $\mathbb{R}^3$ is given by
\begin{equation}
\frac{1}{m_0+Nm}(m_0q_0+mq_1+\cdots+mq_N)
\end{equation}
for all $q_0,q_1,\ldots,q_N\in\mathbb{R}^3$;
\item[3.] there exists $\mu>0$ such that, for any $\psi:\mathbb{R}^{3(1+N)}\to\mathbb{C}$,
\begin{equation}
-\frac{\hbar^2}{2m_0}\Delta_0\psi - \frac{\hbar^2}{2m} \sum_{j=1}^N \Delta_j\psi = -\frac{\hbar^2}{2(m_0+Nm)}\Delta_0(\psi\circ T) - \frac{\hbar^2}{2\mu}\sum_{j=1}^N \Delta_j(\psi\circ T) \ ;
\end{equation}
\item[4.] if $\psi:\mathbb{R}^{3(1+N)}\to\mathbb{C}$ is (anti-)symmetric with respect to exchange of any of its $\mathbb{R}^3$ variables $1$ through $N$, then so is $\psi\circ T^{-1}$;
\end{itemize}
is the one given by (\ref{CMsystem}) with the dimensionless constant $r\in(0,1)$ and the reduced mass $\mu$ given by
\begin{equation}
r = \frac{m_0}{m_0+Nm} \quad , \quad \mu = mr^{1/N} \ .
\label{mu_and_r}
\end{equation}
\label{theorem}
\end{thm}

\begin{rem} When $N=1$, our $T$ recovers the well-known 2-body system of Jacobi coordinates if we choose $\stackrel{(2)}{\pm} = -$. Indeed, in this case we have $\overline{\bm{q}}=q_1$, and equation (\ref{CMsystem}) shows
\begin{equation}
\xi_1 = r^{-1/2}(0 \stackrel{(2)}{\pm}r^{1/2}(q_0-q_1)) = \stackrel{(2)}{\pm}(q_0-q_1) \ .
\end{equation}
Also (\ref{mu_and_r}) implies that $\mu$ is the usual reduced mass (\ref{2D_Jacobi}) for two bodies.
\end{rem}

\vspace*{1 em}

\begin{rem} It is common in statistical problems to consider a more general space of admissible states, called \emph{density matrices}. This is the set $\mathcal{S} = \mathcal{S}(L^2(\mathbb{R}^{3(1+N)}))$ of self-adjoint, positive, trace-class, unit-trace operators acting on $L^2$. In this context, a \emph{pure} state $\psi$ as considered above is represented by the projection operator $|\psi\rangle\langle\psi|$, while a mixture of pure states $\psi_j$ with weights $0<\lambda_j<1$ (such that $\sum_j \lambda_j=1$) gets associated to the operator $\sum_j \lambda_j |\psi_j\rangle\langle\psi_j|$. The expected energy in state $\rho\in\mathcal{S}$ is then given by $Q(\rho) = \mathrm{Tr}[H\rho]$, and the state is called \emph{symmetric} with respect to the nonzero variables when $U_\sigma\rho = \rho$ for all $\sigma\in S^N$, where $U_\sigma$ is the unitary operator $L^2\ni \psi \mapsto \psi\circ\sigma$ (analogously the concept of \emph{anti-symmetry} involves an additional $\mathop{sgn}(\sigma)$).

We can see that the same conditions as in Theorem (\ref{theorem}) will lead to preservation of symmetry in the energy functional as well as space of states in this context. Indeed, when a change of coordinates $T$ is performed on $\mathbb{R}^{3(1+N)}$, the unitary operator $U_T:L^2 \to L^2$, $U_T(\psi)(\bm{q}) = |\det T|^{1/2}\psi(T\bm{q})$ represents the transformation of wavefunctions $\psi$ into the new coordinates. Then a state $|\psi\rangle\langle\psi|$ must become $|U_{T^{-1}}\psi\rangle\langle U_{T^{-1}} \psi|$, and a change of variable in the integral defining the $L^2$ inner product reveals that this is the same as $|\det T|U_{T^{-1}} |\psi\rangle\langle\psi | U_T$. So we have found the expression of a general state $\rho\in\mathcal{S}$ in the new coordinates $T\bm{q}$: it is $|\det T| U_{T^{-1}} \rho U_T$. This immediately implies that preservation of (anti-)symmetry of states is satisfied precisely by the same condition (\ref{T_symm}) as before. Similarly one can consider pure states $|\psi\rangle\langle\psi|$ in order to understand the transformation of the energy functional and find out that (\ref{orthogonality}) is the condition that preserves its symmetry.
\end{rem}

\vspace*{1 em}

To finish writing the transformed Hamiltonian, we need to figure out what the potential part becomes, which requires expressing the $q_j$'s in terms of the $\xi_k$'s. The inverse transformation can be computed from (\ref{TRS}): $T^{-1} = R^{-1} T^t S$, yielding
\begin{equation}
T^{-1} = \begin{pmatrix}
1 & \displaystyle\frac{mr^{\frac{1}{N}}}{m_0}C & \displaystyle\frac{mr^{\frac{1}{N}}}{m_0}C & \displaystyle\frac{mr^{\frac{1}{N}}}{m_0}C & \ldots & \displaystyle\frac{mr^{\frac{1}{N}}}{m_0}C \\ \\
1 & r^{1/N}A & r^{1/N}B & r^{1/N}B & \ldots & r^{1/N}B \\ \\
1 & r^{1/N}B & r^{1/N}A & r^{1/N}B & \ldots & r^{1/N}B \\ \\
1 & r^{1/N}B & r^{1/N}B & r^{1/N}A & \ldots & r^{1/N}B \\
\vdots & \vdots & \vdots & \vdots & \ddots & \vdots \\
1 & r^{1/N}B & r^{1/N}B & r^{1/N}B & \ldots & r^{1/N}A \\
\end{pmatrix}_{(1+N)\times(1+N)} \otimes I_{3\times 3} \ .
\end{equation}
Now plug-in the values of $A,B,C$ given in (\ref{ABC}) to get:
\begin{equation}
\left\{\begin{array}{ccl}
q_0 & = & \xi_0 \stackrel{(2)}{\pm}r^{\frac{1}{2}+\frac{1}{2N}}(r^{-1}-1)\overline{\bm{\xi}} \ , \\
q_j & = & \xi_0 \stackrel{(1)}{\pm}r^{\frac{1}{2N}}\xi_j - \left( \stackrel{(2)}{\pm}r^{\frac{1}{2}+\frac{1}{2N}} \stackrel{(1)}{\pm} r^{\frac{1}{2N}} \right)\overline{\bm{\xi}} \quad , \quad j=1,\ldots,N \ ,
\end{array}\right.
\label{CMinverse}
\end{equation}
where $\overline{\bm{\xi}}$ is defined analogously to how $\overline{\bm{q}}$ was defined in (\ref{empirical_average}). In particular, the relevant pairwise distances for our Hamiltonian and for most physically meaningful others become:
\begin{equation}
\left\{\begin{array}{cclcl}
|q_j-q_0| & = & r^{\frac{1}{2N}}\left| \stackrel{(1)}{\pm}\xi_j - (\stackrel{(1)}{\pm}1\stackrel{(2)}{\pm}r^{-\frac{1}{2}})\overline{\bm{\xi}} \right| & , & j=1,\ldots,N \ , \\
|q_i-q_j| & = & r^{\frac{1}{2N}}|\xi_i-\xi_j| & , & i,j=1,\ldots,N \ .
\end{array}\right.
\label{distances}
\end{equation}
With this, we finally conclude that the potential energy part will transform just like we wished, remaining symmetric with respect to exchanges in the variables other than the zeroth. We conclude that the Hamiltonian that represents the energy intrinsic to the system is given by
\begin{equation}
\tilde{H} = -\frac{\hbar^2}{2mr^{\frac{1}{N}}}\sum_{j=1}^N \Delta_j - \frac{zZ}{r^{\frac{1}{2N}}} \sum_{j=1}^N \frac{1}{\left| \stackrel{(1)}{\pm}\xi_j - (\stackrel{(1)}{\pm} 1 \stackrel{(2)}{\pm} r^{-\frac{1}{2}})\overline{\bm{\xi}} \right|} + \frac{z^2}{r^{\frac{1}{2N}}}\mathop{\sum_{i=1}^N\sum_{j=1}^N}_{i<j} \frac{1}{|\xi_i-\xi_j|} \ .
\label{Htilde}
\end{equation}

\begin{rem} Thinking ahead about what this Hamiltonian has to say, first note that the factors of $r^{1/N}$ and $r^{1/2N}$ in front of each term disappear after a suitable rescaling of the argument of the wavefunction, which aids in understanding how the size of the system in its ground-state scales with $N$. It is also interesting to note that, even after this rescaling, there remains still a dependence on $N$ in the term $\overline{\bm{\xi}}$, which includes a factor $1/N$. Since the arguments used in Hartree and Hartree-Fock theory to study asymptotic properties of the ground-state and equilibrium energy rely heavily on the fact that the Hamiltonian can be written as a sum of individual terms featuring only 1 or 2 variables $\xi_j$ in them (see \cite{roug}), our new transformed problem is not trivial to study. But this is material for future work.
\end{rem}

\section{Symmetric center-of-mass coordinates (classical case)}

Now we explore a different model given by a classical Hamiltonian to see that the same conditions as in theorem (\ref{theorem}) are still the natural ones to require and the conclusions are still mostly the same, in spite of the different nature of the set of states and the form of the kinetic energy part. This time, in order to even be able to talk about symmetry on the set of admissible states, it is necessary to consider them to be statistical distributions of possible phase space configurations (as opposed to the quantum case, where just a single state $\psi$ already comes with a probabilistic interpretation via the Born rule). The proof follows the same ideas as in the quantum case, but applied to different objects that satisfy different properties, and it turns out that the restrictions imposed by this classical context are not enough to warrant uniqueness. Lest the reader be misled into thinking that the Coulomb potential is necessary in the reasoning, we will give the bodies the possibility to interact pairwise through general potential functions - and everything readily generalizes to threefold, fourfold etc. interactions.

Consider a classical-mechanical system evolving in space $\mathbb{R}^3$ consisting of a distinguished body of mass $m_0$ and $N$ identical bodies of mass $m$ such that the potential energy of interaction between the first and any of the others is given by a function $V$, and the one between the identical bodies by a function $W$, both depending symmetrically on the two interacting bodies. The phase space is
\begin{equation}
\mathcal{D} = \mathbb{R}^{3(1+N)}\times\mathbb{R}^{3(1+N)} = \{ (\mathbf{x};\mathbf{p}) = (x_0,x_1,\ldots,x_N;p_0,p_1,\ldots,p_N) \;;\; x_j,p_j\in\mathbb{R}^3 \}
\end{equation}
where each $x_j$ and $p_j$ are the position and momentum of particle $j$ (particle 0 being the distinguished one). The Hamiltonian is the function defined\footnote{More commonly, $H$ is only defined on the subset of $\mathcal{D}$ of the points $(\mathbf{x};\mathbf{p})$ for which no $x_i$ is equal to an $x_j$, $i\neq j$, and the admissible states are measures supported away from such points.} on $\mathcal{D}$ given by
\begin{equation}
H(\mathbf{x};\mathbf{p}) = \frac{1}{2m_0}|p_0|^2 + \frac{1}{2m}\sum_{j=1}^N |p_j|^2 + \sum_{j=1}^N V(x_j,x_0) + \mathop{\sum_{i=1}^N\sum_{j=1}^N}_{i<j} W(\xi_i,\xi_j) \ .
\label{class_Ham}
\end{equation}
The set of states of the system is defined as
\begin{equation}
\mathcal{S} = \text{Set of } \{1,\ldots,N\}\text{-permutation-symmetric Borelian probability measures on } \mathcal{D}
\end{equation}
where the qualification about permutation symmetry means, as one expects for identical bodies, that any $\nu\in\mathcal{S}$ must satisfy
\begin{equation}
\nu(E) = \nu(U_\sigma(E)) \quad , \quad \sigma\in S^N , E\subseteq\mathcal{D} \text{ Borel-measurable},
\end{equation}
where the isomorphism $U_\sigma:\mathcal{D}\to\mathcal{D}$, for $\sigma\in S^N$, is given by
\begin{equation}
U_\sigma(x_0,x_1,\ldots,x_N;p_0,p_1,\ldots,p_N) = (x_0,x_{\sigma(1)},\ldots,x_{\sigma(N)};p_0,p_{\sigma(1)},\ldots,p_{\sigma(N)}) \ .
\end{equation}

If $T:\mathbb{R}^{3(1+N)}\to\mathbb{R}^{3(1+N)}$ is a bijection onto its image (a coordinate change of the position variables), the question arises of how to extend it to the whole $\mathcal{D}$, that is, how to define the physically meaningful transformation $T':\mathbb{R}^{3(1+N)}\to\mathbb{R}^{3(1+N)}$ of the momentum coordinates. Note how this consideration only arises in the present context of classical mechanics, because a fundamental difference between it and quantum theory is that in the latter the stipulation of the space of states does not involve momentum variables.

Assuming for a moment that we have found the correct expression for $T'$, we can identify the transformed phase space $\tilde{D}$, set of admissible states $\tilde{\mathcal{S}}$ and Hamiltonian $\tilde{H}$:
\begin{equation}
    \tilde{\mathcal{D}} = (T\oplus T')(\mathcal{D}) = \{ (T\bm{x};T'\bm{p}) \;;\; (\bm{x};\bm{p})\in\mathcal{D} \} \ ,
\end{equation}
\begin{equation}
    \tilde{S} = \{ (T\oplus T')_\ast\nu \;;\; \nu\in \mathcal{S} \} \ ,
\end{equation}
\begin{equation}
    \tilde{H}(\bm{\xi};\bm{\pi}) = H(T^{-1}\bm{\xi};T'^{-1}\bm{\pi}) \ ,
    \label{new_H}
\end{equation}
where the push-forward probability measure $(T\oplus T')_\ast\nu$ defined by
\begin{equation}
    \left((T\oplus T')_\ast\nu\right) (F) = \nu\left((T\oplus T')^{-1}(F)\right) \qquad , \qquad F\subseteq\tilde{\mathcal{D}} \text{ Borelian}
    \label{new_mu}
\end{equation}
is the state in $\tilde{\mathcal{S}}$ corresponding to a state $\nu\in\mathcal{S}$ (no normalization constant is required). The expression for the Hamiltonian must be given as in (\ref{new_H}) due to the property of push-forward measures that says
\begin{equation}
    \int_{\mathcal{D}} H \mathrm{d}\nu = \int_{\tilde{\mathcal{D}}} (H\circ(T\oplus T')^{-1}) \mathrm{d} (T\oplus T')_\ast\nu \ ,
\end{equation}
that is, the expected value of the energy of the system at the transformed state computed with the transformed Hamiltonian is the same as what it was before the transformation, which is a natural property to desire.

Let us see what kind of conditions are needed for this $T$ to be a symmetric center-of-mass coordinate change for the classical many-body problem at hand. What we called conditions 1 and 2 in theorem (\ref{theorem}) can be stated verbatim here, and in particular they imply that our sought-after $T'$ also satisfies
\begin{equation}
    T' = \tilde{T'}\otimes I_{3\times 3}
\end{equation}
for some isomorphism $\tilde{T'}:\mathbb{R}^{3(1+N)}\to\mathbb{R}^{3(1+N)}$. Condition 4 now asks that, for all $\nu\in\mathcal{S}$, the transformed state (\ref{new_mu}) satisfy $\{1,\ldots,N\}$-symmetry, and there are enough Borelian subsets in Euclidean space to guarantee that this is only possible if the same condition (\ref{T_symm}) as before is valid. Hence our $T$ is of the form (\ref{T_matrix}) (but a similar remark to (\ref{differnet_masses}) applies in case the bodies are not identical and we just wish to avoid cross-terms in the Hamiltonian). Finally condition 3 stipulates that there must exist $\mu>0$ such that
\begin{equation}
\frac{1}{2m_0}|p_0|^2 + \frac{1}{2m}\sum_{j=1}^N|p_j|^2 = \frac{1}{2M_{\mathrm{Tot}}}|\pi_0|^2 + \frac{1}{2\mu}\sum_{j=1}^N|\pi_j|^2
\label{Ham_sym}
\end{equation}
whenever $\bm{\pi} = T'\bm{p}$. By computing $|p_0|^2,|p_j|^2$ in terms of $\pi_i\cdot\pi_j$, we see that the components $T'^{-1}_{ij}$ of $\tilde{T'}^{-1}$, rather than the ones of $T'$ or $\tilde{T'}$ like before, are the ones coming into the computation. The required condition will then be
\begin{equation}
\frac{1}{m_0} T'^{-1}_{0k} T'^{-1}_{0l} + \frac{1}{m}\sum_{j=1}^N T'^{-1}_{jk} T'^{-1}_{jl} = \begin{cases}
1/M_{\mathrm{Tot}} & , \;\;k=l=0 \ , \\
1/\mu & , \;\;k=l>0 \ , \\
0 & , \;\;k\neq l \ .
\end{cases}
\label{new_orthogonality}
\end{equation}
Compare this to (\ref{orthogonality}). The clear difference is that there is no $|\det T'|^{1/2}$ this time, and the subtle difference is the order of the indices $0k$, $0l$, $jk$, $jl$. Using the matrices $R$ and $S$ from (\ref{RS}), equation (\ref{new_orthogonality}) is written as 
\begin{equation}
    \left(\tilde{T'}^{-1}\right)^t R^{-1}\tilde{T'}^{-1} = S^{-1} \ .
    \label{abbrev_ortho}
\end{equation}

Now we can find $T'$ using the expression (\ref{new_H}) of the transformation of $H$ and the desired new form (\ref{Ham_sym}). The Hamilton equations dictate the law of motion of the system in $(\bm{x};\bm{p})$ coordinates and should still hold for the transformed Hamiltonian in $(\bm{\xi}=T\bm{x};\bm{\pi}=T'\bm{p})$ coordinates, the possible curves $(\bm{x}(t);\bm{p}(t))$ that describe the evolution of a point in phase space satisfy
\begin{equation}
    \dot{\bm{x}} = \frac{\partial H}{\partial \bm{p}} \qquad , \qquad \dot{\bm{p}} = -\frac{\partial H}{\partial \bm{x}} \ ,
\end{equation}
and we would like to also have
\begin{equation}
    \dot{\bm{\xi}} = \frac{\partial \tilde{H}}{\partial \bm{\pi}} \qquad , \qquad \dot{\bm{\pi}} = -\frac{\partial \tilde{H}}{\partial \bm{\xi}} \ .
\end{equation}
The zeroth of each of these two systems of equations give
\begin{equation}
    p_0 = m_0\dot{x_0} \; , \; p_j = m\dot{x_j} \; , \; \pi_0 = M_{\mathrm{Tot}}\dot{\xi_0} \; , \; \pi_j = \mu\dot{\xi_j} \; , \; j=1,\ldots,N \ .
\end{equation}
Given $\dot{\bm{\xi}} = T\dot{\bm{x}}$, we conclude
\begin{equation}
    \begin{split}
        \pi_0 &= M_{\mathrm{Tot}}\left( \frac{1}{m_0} T_{00} p_0 + \frac{1}{m}\sum_{k=1}^N T_{0k} p_k \right) \ , \\
        \pi_j &= \quad \;\;\mu\left( \frac{1}{m_0} T_{j0} p_0 + \frac{1}{m}\sum_{k=1}^N T_{jk} p_k \right) \quad , \quad j=1,\ldots,N \ , \\
    \end{split}
\end{equation}
or
\begin{equation}
    \tilde{T'} = S\tilde{T}R^{-1} \ .
    \label{Tprime}
\end{equation}
Together with (\ref{abbrev_ortho}), this implies
\begin{equation}
    \tilde{T'}^{-1} = R(S\tilde{T}R^{-1})^tS^{-1} = \tilde{T}^t \ .
\end{equation}
Plug this into (\ref{new_orthogonality}) to finally see that it actually says precisely the same as (\ref{orthogonality}), except for the $(\det T)^{1/2}$ factor. Hence we may now conclude through the same computations that the transformation (\ref{CMsystem}) satisfies all constraints. However there is nothing that imposes a value for $\mu$ this time. We summarize:

\begin{thm} For given $m_0,m,\mu>0$ and up to two independent choices of $\pm$ signs and relabeling of nonzero indices, the only family of transformations $T:\mathbb{R}^{3(1+N)}\to\mathbb{R}^{3(1+N)}$ (indexed by $N$) satisfying the following:
\begin{itemize}
\item[1.] $T=\tilde{T}\otimes I_{3\times 3}$ for some linear isomorphism $\tilde{T}:\mathbb{R}^{1+N}\to\mathbb{R}^{1+N}$;
\item[2.] the zeroth component of $T(q_0,q_1,\ldots,q_N)$ in $\mathbb{R}^3$ is given by
\begin{equation}
\frac{1}{m_0+Nm}(m_0q_0+mq_1+\cdots+mq_N)
\end{equation}
for all $q_0,q_1,\ldots,q_N\in\mathbb{R}^3$;
\item[3.] if $\bm{\pi} = T'(\bm{p})$, then (\ref{Ham_sym}) is true;
\item[4.] if a probability measure $\nu$ on $\mathbb{R}^{3(1+N)}\times\mathbb{R}^{3(1+N)}$ is symmetric with respect to exchange of any of its $\mathbb{R}^3$ variables $1$ through $N$ in both the first half of its argument (position variables) and the second (momentum variables), then so is the push-forward $(T\oplus T')_\ast\nu$, where $T' = \tilde{T'}\otimes I_{3\times 3}$ is defined by (\ref{Tprime});
\end{itemize}
is the one given by (\ref{T_matrix}) with $A,B,C$ given as in equations (\ref{pm}) and (\ref{C}).
\label{thm_classical}
\end{thm}

\section{Many-species problems}
\label{sec_multi}

Conditions 1 through 4 have analogues that are applicable to problems involving many different groups of identical bodies, which we call \emph{many-species} problems. Here we show that, despite losing uniqueness to the many degrees of freedom afforded by such problems, we can still produce a natural system of center-of-mass coordinates that preserves the symmetries of the Hamiltonian and the permutation symmetry of admissible states with respect to exchange of any two identical bodies. We choose to use quantum-mechanical language again, but it should be clear that the applicability of the result extends to classical physics just like in the above section.

Consider a system containing a distinguished particle\footnote{Since the change of coordinates and subsequent dismissal of the center-of-mass coordinate effectively reduce the number of bodies by one, there should be a body that in a sense gets thrown out of consideration. This doesn't mean that it needs to be the most massive one, the ``nucleus'' or even a different body from all the others; it could as well be within any of the groups $1$ through $n$, but we give it a special notation with the index 0.} of mass $m_0$ at position $q_0\in\mathbb{R}^3$, and $n$ groups of identical particles containing $N_1,\ldots,N_n$ particles. We must assume that each $N_i$ is at least $2$. Denote by $1+N$ the total number of particles:
\begin{equation}
1+N = 1+N_1+\cdots+N_n \ .
\end{equation}
Suppose that the particles belonging to group $i$ all have mass $m_i$ and are located at $q^{(i)}_1,\ldots,q^{(i)}_{N_i}$. Let the energy of interaction between the zeroth particle and a particle of group $i$ be given by a function $V_i$ depending symmetrically on their positions; let the energy of interaction between a particle of group $i$ and another of group $j$ (possibly $i=j$) be given by a function $W_{i,j}$ depending symmetrically on their positions. The Hamiltonian
\begin{multline}
    H = -\frac{\hbar^2}{2m_0}\Delta_{q_0} - \frac{\hbar^2}{2}\sum_{i=1}^n \frac{1}{m_i}\sum_{k=1}^{N_i} \Delta_{q^{(i)}_{k}} + \sum_{i=1}^n\sum_{k=1}^{N_i} V_i(q_0,q^{(i)}_k) + \\ + \mathop{\sum_{i=1}^n\sum_{j=1}^n}_{i\leq j} \mathop{\sum_{k=1}^{N_i}\sum_{l=1}^{N_j}}_{k<l} W_{i,j}(q^{(i)}_k,q^{(j)}_l)
    \label{multi_H}
\end{multline}
(in self-explanatory notation for the Laplacians) is defined on a suitable subset of the space of admissible wavefunctions, which are those twice-differentiable $L^2$ functions of $\mathbb{R}^{3(1+N)}$ that are (anti-)symmetric with respect to exchange of any two variables $q^{(i)}_k$ and $q^{(i)}_l$ of the same group.

The change-of-coordinates maps that we seek are in the form $T=\tilde{T}\otimes I_{3\times3}$, for $\tilde{T}:\mathbb{R}^{1+N}\to \mathbb{R}^{1+N}$ a linear map whose matrix has zeroth row given by
\begin{equation}
    (T_{0j}) = \left( \frac{m_0}{M_{\mathrm{Tot}}} , \left[\frac{m_1}{M_{\mathrm{Tot}}}\right]_{N_1} , \ldots , \left[\frac{m_n}{M_{\mathrm{Tot}}}\right]_{N_n} \right)
\end{equation}
where $M_{\mathrm{Tot}} = m_0 + \sum_i N_im_i$. Here the notation $[ x ]_k$ represents a vector $(x,\ldots,x)$ with $k$ components. But it's best to label the rows and columns of $\tilde{T}$ with the symbols
\begin{equation}
    0,1^{(1)},\ldots,N_1^{(1)},1^{(2)},\ldots,N_2^{(2)},\ldots,1^{(n)},\ldots,N_n^{(n)}
\end{equation}
in this order. For example, the entry in the row corresponding to the third particle of group 5 and the column corresponding to the second-to-last particle of group 1 would then be $T_{3^{(5)} , (N_1-1)^{(1)}}$.

Due to permutation symmetry of states, a property analogous to (\ref{T_symm}) must hold, which can be stated as follows: for fixed $i=1,\ldots,n$, to each permutation $\sigma$ of $\{1,\ldots,N_i\}$ there must correspond a permutation $\pi$ comprised of permutations within each group $1,\ldots,N$ (not necessarily only group $i$), such that $\tilde{T}$ remains unchanged under swapping of its rows according to $\sigma$ followed by swapping of its columns according to $\pi$. We shall not attempt to classify all possible ways to construct a $\pi$ for each $\sigma$ if $\pi$ is allowed to permute variables of many groups; instead let us consider that $\pi$ must only act on group $i$. In the same way, to each permutation $\pi$ of columns within a group corresponds a permutation $\sigma$ of rows of that group such that performing $\pi$ followed by $\sigma$ on $\tilde{T}$ leaves it unchanged. Then this implies:
\begin{itemize}
    \item that the zeroth column must be of the form
    \begin{equation}
        (T_{k0})_k = \left( \frac{m_0}{M_{\mathrm{Tot}}} , \left[ C_1 \right]_{N_1} , \ldots , \left[ C_n \right]_{N_n} \right)^t
    \end{equation}
    for some numbers $C_1,\ldots,C_n$ (consider what happens when swapping any two rows of group $i$);
    \item that each of the $n$ square blocks on the main diagonal must be of the form
    \begin{equation}
        (T_{kl})_{k,l = 1^{(i)},\ldots,N_i^{(i)}} = \begin{bmatrix} A_i & B_i & B_i & \cdots & B_i \\ B_i & A_i & B_i & \cdots & B_i \\ B_i & B_i & A_i & \cdots & B_i \\ \vdots & \vdots & \vdots & \ddots & \vdots \\ B_i & B_i & B_i & \cdots & A_i \end{bmatrix}
    \end{equation}
    for some numbers $A_i,B_i$ (that is, after a permutation of indices $1^{(i)},\ldots,N_i^{(i)}$; this is just like the proof in Section 3);
    \item that, given any $i\neq j$, the rectangular off-diagonal block whose rows are in group $i$ and columns are in group $j$ must have all elements equal to the same number $X_{ij}$ (consider first what happens when swapping rows of group $i$, then also what happens when swapping columns of group $j$).
\end{itemize}

Finally, to prevent cross-terms in the kinetic energy and preserve its symmetries, we require the same condition as in (\ref{TRS}):
\begin{equation}
\tilde{T}R^{-1}\tilde{T}^t = S^{-1}
\label{ortho_multiple}
\end{equation}
where now
\begin{equation}
    R = \mathrm{diag} \left( m_0 , [m_1]_{N_1} , \ldots , [m_n]_{N_m} \right) \quad , \quad S = \mathrm{diag} \left( M_{\mathrm{Tot}} , [\mu_1]_{N_1} , \ldots , [\mu_n]_{N_m} \right)
\end{equation}
for some reduced masses $\mu_1,\ldots,\mu_n>0$. The condition $|\det \tilde{T}|=1$, present in the quantum context but not the classical one, implies that we must impose
\begin{equation}
    m_0 m_1 \cdots m_n = M_{\mathrm{Tot}} \mu_1 \cdots \mu_n \ .
    \label{reduced_masses_constraint}
\end{equation}

Now it becomes more convenient to normalize the elements of $\tilde{T}$ by considering the matrix
\begin{equation}
    U = S^{1/2} \tilde{T} R^{-1/2} \ ,
    \label{U_from_T_tilde}
\end{equation}
which, according to (\ref{ortho_multiple}), must be orthogonal. It is obtained from $\tilde{T}$ by multiplying row $0$ by $\sqrt{M_{\mathrm{Tot}}}$ and rows $k^{(i)}$ by $\sqrt{\mu_i}$ and dividing column $0$ by $\sqrt{m_0}$ and columns $k^{(i)}$ by $\sqrt{m_i}$. Its zeroth row is then determined:
\begin{equation}
    (U_{0j}) = \left( \sqrt{\frac{m_0}{M_{\mathrm{Tot}}}} , \left[\sqrt{\frac{m_1}{M_{\mathrm{Tot}}}} \ \right]_{N_1} , \ldots , \left[\sqrt{\frac{m_n}{M_{\mathrm{Tot}}}} \ \right]_{N_n} \right) \ ,
    \label{nus}
\end{equation}
and is already normalized to $1$ in Euclidean norm. Let us abbreviate it by using the symbols
\begin{equation}
    (U_{0j}) = \left( \nu_0 , [\nu_1]_{N_1} , \ldots , [\nu_n]_{N_n} \right) \ ,
    \label{first_row_of_U}
\end{equation}
(all of them are determined by the data of the problem) and denote the other elements of $U$ with lowercase letters $a_i,b_i,c_i,x_{ij}$ in the locations corresponding to $A_i,B_i,C_i,X_{ij}$ in $\tilde{T}$. We have thus reduced the question to the following: given real constants $\nu_0,\nu_1,\ldots,\nu_n$ satisfying
\begin{equation}
    \nu_0^2 + N_1\nu_1^2 + \cdots + N_n\nu_n^2 = 1 \ ,
    \label{sum_squares}
\end{equation}
can one find an orthogonal matrix $U$ in the following format?
\begin{equation}
    U = \left[\begin{array}{c|c|c|c|c|c}
    \nu_0 & \nu_1 \; \nu_1 \; \cdots \; \nu_1 & \nu_2 \; \nu_2 \; \cdots \; \nu_2 & \nu_3 \; \nu_3 \; \cdots \; \nu_3 & \cdots & \nu_n \; \nu_n \; \cdots \; \nu_n \\ \hline
    {\begin{array}{c} c_1 \\ \vdots \\ c_1 \end{array} } & \mathcal{U}_{11} & \mathcal{U}_{12} & \mathcal{U}_{13} & \cdots & \mathcal{U}_{1n} \\ \hline
    {\begin{array}{c} c_2 \\ \vdots \\ c_2 \end{array} } & \mathcal{U}_{21} & \mathcal{U}_{22} & \mathcal{U}_{23} & \cdots & \mathcal{U}_{2n} \\ \hline
    \vdots & \vdots & \vdots & \vdots & \ddots & \vdots \\ \hline
    {\begin{array}{c} c_n \\ \vdots \\ c_n \end{array} } & \mathcal{U}_{n1} & \mathcal{U}_{n2} & \mathcal{U}_{n3} & \cdots & \mathcal{U}_{nn} \\
    \end{array}\right],
    \label{form_of_U}
\end{equation}
where the blocks $(\mathcal{U}_{ij})_{N_i\times N_j}$ are of the following form:
\begin{equation}
    \mathcal{U}_{ii} = \begin{bmatrix} a_i & b_i & b_i & \cdots & b_i \\ b_i & a_i & b_i & \cdots & b_i \\ b_i & b_i & a_i & \cdots & b_i \\ \vdots & \vdots & \vdots & \ddots & \vdots \\ b_i & b_i & b_i & \cdots & a_i \end{bmatrix} \qquad , \qquad \mathcal{U}_{ij} = \begin{bmatrix} x_{ij} & x_{ij} & x_{ij} & \cdots & x_{ij} \\ x_{ij} & x_{ij} & x_{ij} & \cdots & x_{ij} \\ x_{ij} & x_{ij} & x_{ij} & \cdots & x_{ij} \\ \vdots & \vdots & \vdots & \ddots & \vdots \\ x_{ij} & x_{ij} & x_{ij} & \cdots & x_{ij} \end{bmatrix} \; (i\neq j) \ .
    \label{form_of_blocks}
\end{equation}
Now there are $n$ equations imposing norm $1$ for each row and $n$ equations imposing orthogonality of each row with the zeroth (within each group these are all the same), plus $n$ equations imposing orthogonality of different rows within the same group, plus $n(n-1)/2$ equations imposing orthogonality of rows in different groups, for a total of $n(n+5)/2$ equations. Meanwhile, there are $n$ variables in the zeroth column of $U$, $2$ in each of its $n$ diagonal blocks, 1 for each one of the $n(n-1)/2$ rectangular blocks above the diagonal, and the same for the blocks below, for a total of $n^2+2n$ variables. The number of degrees of freedom is then computed to be $n(n-1)/2$, exactly the same as the number of blocks above or below the diagonal. Hence there won't be a unique solution, but the numbers suggest we might still be able to solve all these equations by also imposing $n(n-1)/2$ conditions; let us impose $x_{ij} = x_{ji}$ for all $i\neq j$. Then the big square block of $U$ consisting of all rows and columns except the zeroth is symmetric. Now consider the equations that impose norm one for the nonzeroth rows and columns:
\begin{equation}
    \begin{split}
        c_i^2 + a_i^2 + (N_i-1)b_i^2 + \sum_{j\neq i} x_{ij}^2 &= 1 \quad , \quad i=1,\ldots,n \ , \\
        \nu_i^2 + a_i^2 + (N_i-1)b_i^2 + \sum_{j\neq i} x_{ji}^2 &= 1 \quad , \quad i=1,\ldots,n \ . \\
    \end{split}
    \label{c_and_nu}
\end{equation}
With our choice $x_{ij}=x_{ji}$, we see that $c_i = \pm \nu_i$. We choose
\begin{equation}
    c_i = \nu_i \quad , \quad i=1,\ldots,n
    \label{c_i}
\end{equation}
to make $U$ a symmetric matrix. Then it is diagonalizable and admits a basis of orthogonal eigenvectors: there exists an orthogonal matrix $\mathcal{O}$ and a diagonal matrix $D$ such that
\begin{equation}
    U = \mathcal{O} D \mathcal{O}^t \ .
    \label{diagonalize}
\end{equation}
Therefore $U$ is going to be orthogonal if and only if
\begin{equation}
    I = UU^t = U^2 = \mathcal{O}D^2\mathcal{O}^t \; \Longleftrightarrow \; D^2 = I \ ,
\end{equation}
if and only if its eigenvalues are all $\pm 1$. Of course at least one eigenvalue 1 and one $-1$ need to be present, otherwise $U$ would be diagonal according to (\ref{diagonalize}).

Just like what happened in the one-species problem, each $a_i$ depends in a simple way on $b_i$: subtracting from the first equation in (\ref{c_and_nu}) the equation that says that any two different rows in group $i$ are orthogonal, we have
\begin{equation}
c_i^2 + 2a_ib_i + (N_i-2)b_i^2 + \sum_{j\neq i} x_{ji}^2 = 0 \ ,
\end{equation}
so that
\begin{equation}
a_i^2 - 2a_ib_i + b_i^2 = 1 \; \Longleftrightarrow \; a_i = b_i \pm 1 \ .
\end{equation}
We will choose
\begin{equation}
a_i = b_j-1 \quad \text{for all } i \ .
\label{a_i}
\end{equation}
With this, it is possible to choose values for each $b_i$ and $x_{ij}$ that force $U$ to have eigenvalue $-1$ with very high multiplicity: choose
\begin{equation}
    b_i = \rho \nu_i^2 \quad , \quad x_{ij} = \rho \nu_i\nu_j
    \label{b_i}
\end{equation}
for some $\rho>0$ to be determined shortly. Then $U+I$ has column $k^{(i)}$ equal to
\begin{equation}
(U_{l^{(j)},k^{(i)}})_{l^{(j)}=0,\ldots,N} = \left( \nu_i , [\rho \nu_1\nu_i]_{N_1} , \ldots , [\rho\nu_n\nu_i]_{N_n} \right)^t.
\end{equation}
This is a multiple of the vector $(1,[\rho\nu_1]_{N_1},\ldots,[\rho\nu_n]_{N_n})^t$, which is independent of $i$ or $k$. So all nonzeroth columns of $U+I$ are multiples of each other, giving this matrix a rank of at most $2$, and giving $-1$ a multiplicity of at most $N-1$. With the further choice
\begin{equation}
    \rho = \frac{1}{1+\nu_0} \ ,
    \label{rho}
\end{equation}
the zeroth column is also a multiple of that same vector, and $-1$ will have multiplicity $N$ (we remark that algebraic and geometric multiplicity are the same in this case since $U$ is diagonalizable).

Hence $U$ has just one other eigenvalue, $\lambda$, which we must check is equal to $1$. For that purpose, note that a basis for the $-1$ eigenspace is given by vectors $\{w_k^{(i)} \;;\; i=1,\ldots,n \;,\; k=1,\ldots,N_i \}$, each having only two nonzero components: the zeroth entry equal to $1$ and the $k^{(i)\text{th}}$ entry equal to $-1/\rho\nu_i$. Indeed, these are clearly $N$ independent vectors and
\begin{equation}
    (U+I) w_k^{(i)} = \left( (1+\nu_0) - \frac{\nu_i}{\rho\nu_i} , \left[ \nu_1 - \frac{\rho\nu_1\nu_i}{\rho\nu_i} \right]_{N^1} , \ldots , \left[\nu_n - \frac{\rho\nu_n\nu_i}{\rho\nu_i} \right]_{N^n} \right)^t = 0 \ .
\end{equation}
A vector orthogonal to all the $w_k^{(i)}$ is easily constructed:
\begin{equation}
    w_0 = (1 , [\rho\nu_1]_{N_1} , \ldots , [\rho\nu_n]_{N_n})^t \ .
\end{equation}
Because $U$ is symmetric, eigenvectors corresponding to different eigenvalues are orthogonal, so the eigenspace corresponding to $\lambda$ must be spanned by $w_0$. Now simply note that the zeroth coordinate of $Uw_0$ is
\begin{equation}
    \nu_0 + \sum_{i=1}^n \sum_{k=1}^{N_i} \nu_i\rho\nu_i = \nu_0 + \rho\sum_{i=1}^n N_i\nu_i^2 \ ,
\end{equation}
which due to (\ref{sum_squares}) becomes just
\begin{equation}
    \nu_0 + \rho (1-\nu_0^2) = \nu_0 + \frac{1-\nu_0^2}{1+\nu_0} = \nu_0 + (1-\nu_0) = 1 \ ,
\end{equation}
the same as the zeroth coordinate of $w_0$ itself. Hence $\lambda=1$, and $U$ is orthogonal as needed.

Going back through (\ref{rho}), (\ref{b_i}), (\ref{a_i}), (\ref{c_i}), (\ref{form_of_blocks}), (\ref{form_of_U}), (\ref{first_row_of_U}), (\ref{nus}) and (\ref{U_from_T_tilde}), we can finally write our change of coordinates. Letting $\bm{\xi} = T\bm{q}$, we already know that $\xi_0$ is the center-of-mass of the system, and for the rest we can compute:
\begin{equation}
    \begin{split}
        \xi_k^{(i)} &= \sqrt{\frac{m_0}{\mu_i}}\nu_i q_0 - \sqrt{\frac{m_i}{\mu_i}}q_k^{(i)} + \frac{1}{1+\nu_0}\sum_{j=1}^n \nu_i\nu_j \sum_{l=1}^{N_j} \sqrt{\frac{m_j}{\mu_i}}q_l^{(j)} \\
        &= \frac{1}{\sqrt{\mu_i}} \left( \sqrt{\frac{m_0m_i}{M_{\mathrm{Tot}}}}q_0 - \sqrt{m_i}q_k^{(i)} + \frac{1}{1+\sqrt{\frac{m_0}{M_{\mathrm{Tot}}}}}\sum_{j=1}^n \frac{m_j\sqrt{m_i}}{M_{\mathrm{Tot}}}\sum_{l=1}^{N_j} q_l^{(j)} \right) \\
        &= \sqrt{\frac{m_i}{\mu_i}} \left( \sqrt{\frac{m_0}{M_{\mathrm{Tot}}}}q_0 - q_k^{(i)} + \frac{1}{M_{\mathrm{Tot}}+\sqrt{M_{\mathrm{Tot}}m_0}}\sum_{j=1}^n m_j \sum_{l=1}^{N_j} q_l^{(j)} \right) \\
        &= \sqrt{\frac{m_i}{\mu_i}} \left( \sqrt{\frac{m_0}{M_{\mathrm{Tot}}}}q_0 - q_k^{(i)} + \left( 1-\sqrt{\frac{m_0}{M_{\mathrm{Tot}}}} \right)\frac{1}{M_{\mathrm{Tot}}-m_0}\sum_{j=1}^n m_j \sum_{l=1}^{N_j} q_l^{(j)} \right) \\
    \end{split}
\end{equation}
where we deliberately arranged for the center-of-mass of all but the zeroth particle to appear. Wrapping it all up in a theorem:

\begin{thm} Given positive integers $n\geq 1$, $N_1,\ldots,N_n\geq 2$ and positive real numbers $m_0,m_1,\ldots,m_n,\mu_1,\ldots,\mu_n$, let $M_{\mathrm{Tot}}=m_0+N_1m_1+\ldots+N_nm_m$. Then a possible linear transformation
\begin{equation}
    \begin{split}
    T = \tilde{T} \otimes I_{3\times 3} : \mathbb{R}^{3(1+N_1+\ldots+N_n)} &\to \mathbb{R}^{3(1+N_1+\ldots+N_n)} \\ (q_0,(q^{(1)}_k)_{k=1,\ldots,N_1},\ldots,(q^{(n)}_k)_{k=1,\ldots,N_n})^t &\mapsto (\xi_0,(\xi^{(1)}_k)_{k=1,\ldots,N_1},\ldots,(\xi^{(n)}_k)_{k=1,\ldots,N_n})^t
    \end{split}
\end{equation}
that preserves the permutation symmetries and structure of the many-species Hamiltonian (\ref{multi_H}), as well as the symmetry of the admissible states of the $(1+N_1+\cdots+N_n)$-body problem associated with it, is the following:
\begin{equation}
    \left\{\begin{array}{ccl}
        \xi_0 &=& \displaystyle\frac{1}{M_{\mathrm{Tot}}}\left( m_0q_0 + \sum_{i=1}^n m_i \sum_{k=1}^{N_i} q_k^{(i)} \right) \\
        \xi_k^{(i)} &=& \sqrt{\frac{m_i}{\mu_i}} \left( \sqrt{\frac{m_0}{M_{\mathrm{Tot}}}}(q_0-\overline{\bm{q}}) + \overline{\bm{q}}-q_k^{(i)} \right)
    \end{array}\right.
\end{equation}
where we used this abbreviation:
\begin{equation}
    \overline{\bm{q}} = \frac{1}{M_{\mathrm{Tot}}-m_0}\sum_{j=1}^n m_j \sum_{l=1}^{N_j} q_l^{(j)}.
\end{equation}
\label{thm_multi}
\end{thm}

The inverse transformation is obtained from (\ref{U_from_T_tilde}):
\begin{equation}
    \tilde{T}^{-1} = (S^{-1/2} U R^{1/2})^{-1} = R^{-1/2}US^{1/2} \ .
\end{equation}
We compute:
\begin{equation}
        q_0 = \sqrt{\frac{M_{\mathrm{Tot}}}{m_0}} \nu_0 \xi_0 + \sum_{j=1}^n \sum_{l=1}^{N_j} \sqrt{\frac{\mu_j}{m_0}} \nu_j \xi_l^{(j)} = \xi_0 + \sum_{j=1}^n \sqrt{\frac{\mu_j m_j}{M_{\mathrm{Tot}}m_0}} \sum_{l=1}^{N_j} \xi_l^{(j)}
\end{equation}
and
\begin{equation}
    \begin{split}
        q_k^{(i)} &= \sqrt{\frac{M_{\mathrm{Tot}}}{m_i}}\nu_i \xi_0 - \sqrt{\frac{\mu_i}{m_i}}\xi_k^{(i)} + \frac{1}{1+\nu_0}\sum_{j=1}^n \nu_i\nu_j \sum_{l=1}^{N_j} \sqrt{\frac{\mu_j}{m_i}}\xi_l^{(j)} \\
        &= \frac{1}{\sqrt{m_i}} \left( \sqrt{m_i} \xi_0 - \sqrt{\mu_i}\xi_k^{(i)} + \frac{1}{1+\sqrt{\frac{m_0}{M_{\mathrm{Tot}}}}} \sum_{j=1}^n \frac{\sqrt{m_im_j\mu_j}{}}{M_{\mathrm{Tot}}}\sum_{k=1}^{N_j} \xi_l^{(j)} \right) \\
        &= \xi_0 - \sqrt{\frac{\mu_i}{m_i}}\xi_k^{(i)} + \frac{1}{M_{\mathrm{Tot}}+\sqrt{M_{\mathrm{Tot}}m_0}} \sum_{j=1}^n \sqrt{m_j\mu_j}\sum_{k=1}^{N_j} \xi_l^{(j)} \ .
    \end{split}
\end{equation}
Interestingly, unlike what happened in the one-species problem, the analogous quantity to $\overline{\bm{q}}$, which would be
\begin{equation}
    \overline{\bm{\xi}} = \frac{1}{N_1\mu_1 + \cdots + N_n\mu_n}\sum_{j=1}^n \mu_j \sum_{l=1}^{N_j} \xi_l^{(j)} \ ,
\end{equation}
does not appear in these formulas, unless we specifically choose $\mu_i = m_i$ for all $i$.

\section{Acknowledgments}

The author would like to thank his PhD advisor Michael Kiessling and his co-advisor Shadi Tahvildar-Zadeh for valuable discussions, and his fellow graduate students Keith Frankston and Justin Semonsen for conversations about linear algebra that turned out to be pertinent here. Thanks also go to Nicolas Rougerie for his helpful suggestions of references, including his set of lecture notes \cite{roug}, and to Sheldon Goldstein for his insightful comments about Jacobi coordinates and statistical mechanics.

\end{document}